\documentclass{aa} 
\usepackage{graphicx}
 
\begin{document}

\thesaurus{08        %
           ( 9.16.2  
             9.11.1  
             9.01.2  
                   }    

\title{The tetra-lobed planetary nebula NGC 1501}
\author{F.\ Sabbadin\inst{1} \and  S. Benetti\inst{2} \and 
E. Cappellaro\inst{1} \and M. Turatto\inst{1}}
\institute{Osservatorio Astronomico di Padova, vicolo dell'Osservatorio 5,
I-35122 Padova, Italy \and Telescopio 
Nazionale Galileo, Aptdo. Correos 565, E-38700 Santa Cruz de la Palma, 
Canary Island, Spain}

\offprints{F. Sabbadin}

\date{Received ................; accepted ................}

\maketitle

\begin{abstract}

Direct imagery and long-slit, spatially resolved echellograms of the high 
excitation planetary nebula NGC 1501 allowed us to study in detail the 
expansion 
velocity field, the physical conditions (electron temperature, electron 
density, ionization) and the spatial distribution of the nebular gas.

An electron temperature of 11500 K and a turbulence of 18 km s$^{-1}$ are
derived by comparing the H$\alpha$ 
and [OIII] emission line profiles, but large, small scale fluctuations of 
both these 
quantities are present in the ionized gas.

The radial density distribution shows external peaks up to 1400 
cm$^{-3}$; they have steep outwards profiles and extended inwards tails 
probably originated by Rayleigh-Taylor instability and winds interaction.
 
The complexity of the expanding motions indicates that 
the main part of NGC 1501 is a thin ellipsoid of moderate 
ellipticity, but the presence of a pair of large lobes along both 
the major and 
the intermediate axes and of a multitude of smaller bumps spread on the 
whole nebular surface, makes the general 3-D structure of NGC 1501 like a 
boiling, tetra-lobed shell.

This peculiar morphology can be qualitatively explained in terms of 
interaction of the slow nebular material with the intense and fast wind from 
the WC4/OVI central star.
 
\end{abstract}


\section{Introduction}

\begin{figure}
\resizebox{\hsize}{!}{\includegraphics*{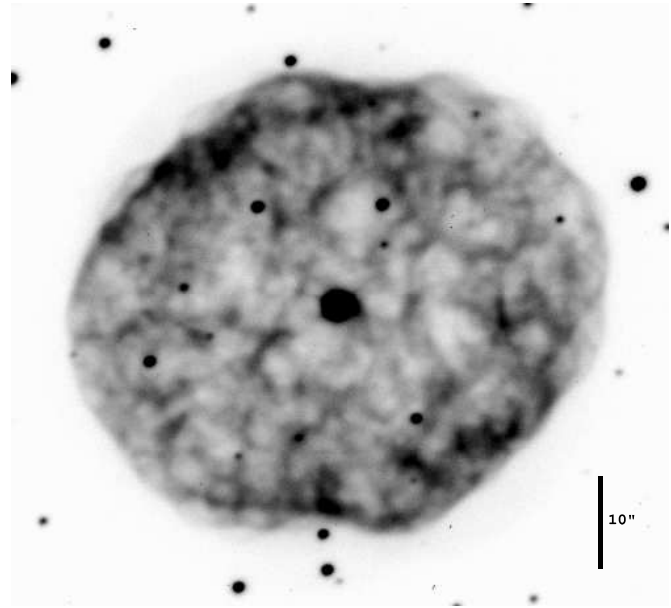}}

\caption{Broad-band R image of NGC 1501
(exposure time 600s, seeing 0.76$\arcsec$) obtained with the 
3.58m Italian National Telescope (TNG); North is up and East is on the 
left. At higher contrast a very faint, diffuse 
and roundish envelope appears, extending up to 34$\arcsec$ from the 
central star.
}\label{figura1}
\end{figure}

The little studied, high excitation planetary nebula (PN) NGC 1501 
(PNG144.5+06.5,  Acker et al., 1992) is ``a very irregular and  patchy elliptical
disk, about 56''x48'' in P.A.=98$^o$. The periphery shows  traces of a broken
ring formation; the brightest portions are the edges at the  ends of the  minor
axis. Relatively faint'' (Curtis, 1918).

At least a dozen of distance estimates of this nebula are  contained in the
literature.  They span in the range 0.9 Kpc (Amnuel et al., 1984; statistical 
distance based on the surface radio-brightness) to 2.0 Kpc (Acker, 1978; 
individual distance based on the average extinction in the galactic disk) and 
roughly peak around 1.3 Kpc (the value we will adopt in the present  paper). 

A kinematical study by Sabbadin and Hamzaoglu (1982 a) suggests that NGC 1501 
is a prolate spheroid of moderate ellipticity.

The spectral type of the exciting star is WC4/ OVI (Aller, 1976,  Tylenda et
al., 1993, Gorny and Stasinska, 1995). The star has a high  temperature, between
$\log T_{\rm Z(HeII)} = 4.91 - 4.98 \mbox{ K}$ (Sabbadin, 1986, Stanghellini  et
al., 1994)  and $\log T_{\rm model~atmosphere}= 5.13 \mbox{ K}$ (Koesterke and
Hamann,  1997a). It is loosing mass at a rate of  $5.2 \times 10^{-7} \mbox{ M}\sun 
\mbox{ yr}^{-1}$ (Koesterke and Hamann, 1997a)  and terminal  wind
velocities of 1800 or 3300 $\mbox{ km\,s}^{-1}$ (Koesterke  and Hamann, 1997a
and Feibelman, 1998, respectively).

It is one of few PNe nuclei showing nonradial g-mode pulsations (Bond et  al.,
1993, 1996, and Ciardullo and Bond, 1996). Following the "born-again  scenario" 
proposed by Iben et al. (1983; see also Bl\"ocker, 1985 and Herwig et al., 
1999), these extremely hot, hydrogen deficient stars suffer a late thermal  pulse
(ejecting the hydrogen rich layers and exposing the naked C/O core)  after they
reached the White  Dwarf cooling sequence and are possible precursors of the
short-period GW Vir  (PG 1159-035) pulsating White Dwarfs.  Hamann (1997) and
Koesterke and Hamann (1997b) suggest the evolutionary sequence [WC-late]
$\rightarrow$ [WC-early] $\rightarrow$ [WC-PG1159] $\rightarrow$\\ PG1159. 

\begin{figure}
\resizebox{\hsize}{!}{\includegraphics*{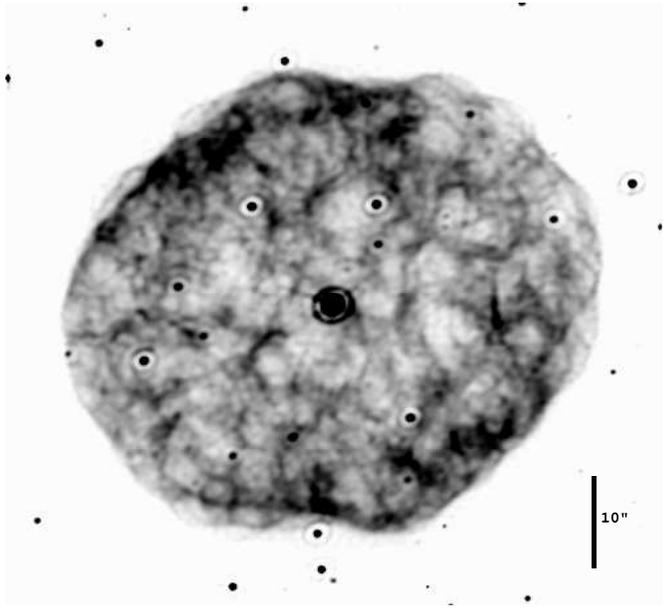}}
\caption{The same R frame of Fig. 1 after the application of a soft 
Lucy-Richardson restoration (point spread function=seeing, 10 iterations), 
showing in great detail the tangled H$\alpha$ structure of NGC 1501. 
The haloes around stars projected onto the nebula are artifacts due to the 
deconvolution.}\label{figura2}
\end{figure}

Recently we started a long-term observing program aimed to investigate 
the physical conditions and the spatial structure of selected PNe through 
direct imagery and 
long-slit Echelle spectra.
This article concerns NGC 1501 and its plane is as follows: in Section 2 we 
present the 
observational material; Section 3 contains  
the expansion velocity field; the physical conditions (electron temperature,
turbulence  
and electron density) of the ionized gas are analysed in Section 4 
and the tomographic maps in 
Section 5; in Section 6 we describe and discuss the resulting spatial model; 
conclusions are drawn in Section 7.

\section{Observational material}

\subsection{Imagery}

The best imagery of NGC 1501 available in the literature goes back 
to Minkowski (1968), who 
obtained H$\alpha$ and [OIII] interference filter plates with the Hale 
Telescope at Mount Palomar, revealing the intricate nebular structure  
(but see also Chu et al., 1987, and Manchado et al., 1996a). 

The superior spatial resolution of HST hasn't yet been exploited for  NGC 1501;
at the moment, only  four broad-band, quite underexposed frames of the  nebula
(taken by Howard Bond on September 7,  1995, during a snapshot survey for
companions of PNe nuclei) are contained  in the HST archives; because of the 
low  surface brightness of the emitting gas, they little add to Minkowski's 
H$\alpha$ plate published in the frontispiece of the IAU  Symposium N. 34
proceedings.

On September 1999, a broad-band R image of NGC 1501 (exposure time 600s,  seeing
0.76'') was  obtained with the Optical Imager Galileo (OIG) mounted on the 3.58m 
Italian  National Telescope (TNG, Roque de los Muchachos, La Palma, Canary
Islands) and equipped with a mosaic of two thinned, back-illuminated  EEV42-80
CCDs with 2048x4096 pixels each (pixel size=13.5 microns; pixel  scale in 2x2
binned mode=0.144 $\arcsec$ pix$^{-1}$).  The CCD data array was processed in the
usual way using standard  IRAF packages.

The appearance of NGC 1501 in the R band, shown in Figure 1, is almost 
entirely due to 
H$\alpha$ 
emission; the contamination of [NII] ($\lambda$$\lambda$6548 and 6584 
\AA\/) and of HeI ($\lambda$5876 \AA\/) amounts to 4\%\ and 3\%\/, 
respectively; 
other fainter lines contribute for less than 2\%\ (see Kaler, 1976, 
Stanghellini et al., 1994, and the next sub-section).

Our R image reveals the presence of a diffuse, roundish envelope 
framing the main nebular body and extending up to a distance of 34'' from the 
central star (this halo is too faint to be reproduced in Fig. 1).

The application of a soft Lucy-Richardson restoration (point spread 
function=seeing, 10 iterations; Figure 2) to the original R frame highlights 
the 
bubbly structure of NGC 1501, rich of filaments and condensations (``bearing 
a resemblance to the convolutions of the brain'' is the suggestive description 
given by Francis G. Pease in 1917).

\subsection{High resolution spectra}

On December 1998, $\lambda$$\lambda$4500-8000 \AA\/ spatially resolved, 
long-slit spectra 
of NGC 1501 (+ flat fields + Th-Ar calibrations + comparison star 
spectra) were secured with the
Echelle spectrograph (Sabbadin and Hamzaoglu, 1981, 1982 b) attached to the
Cassegrain focus of the 182cm Asiago telescope and 
equipped with a Thompson 1024x1024 pixels CCD. 

We selected four position angles: 100$^o$ 
(apparent major axis, after Curtis, 1918), 10$^o$ (perpendicular to the 
apparent major axis),
55$^o$ and 145$^o$ (intermediate positions). All spectra were centred on the 
exciting star; we used the stellar continuum as a position marker and to 
correct the data for seeing and guiding uncertainties. The slit-width
was 0.200 mm (2.5$\arcsec$ on the sky), corresponding to a spectral
resolution of 13.5 km s$^{-1}$ (1.5 pixel).  
The calibration in wavelength and flux was performed in the standard way using 
the IRAF Echelle packages.

H$\alpha$ and $\lambda$5007 \AA\/ of [OIII], the dominant emissions in our 
spectra, present the same structure in every 
detail (but not in the line width: as normally observed, the first is broader 
than the second, because of the larger thermal motions).
Besides H$\alpha$ and $\lambda$5007 \AA\/ (and the correlated lines 
H$\beta$ and $\lambda$4959 \AA\/), only a few, faint emissions were detected, 
due to the low surface brightness of the emitting gas; they all mimic the 
H$\alpha$ and [OIII] structure. The line intensities,
integrated over the whole nebula and the entire velocity profile, are listed 
in Table 1; only relative fluxes are reported here, since the observational 
conditions (the sky was stable only at intervals) prevent the accurate 
calibration into absolute fluxes.

The interstellar extinction was derived by comparing the observed  ${\rm
H}\alpha/{\rm H}\beta$ intensity ratio with the dereddened value given by 
Brocklehurst  (1971) for $T{\rm e}=10^4 \mbox{ K}$ and $N{\rm e}=10^4 \mbox{
cm}^{-3}$; we obtain $c({\rm H}\beta)= 1.05 \pm 0.10$, in agreement with the
values of 1.1 and 0.96 reported by  Kaler (1976) and Stanghellini et al. (1994),
respectively.

\begin {table}
\caption{Integrated line intensities}
\begin{flushleft}
\begin{tabular}{lcrcc}
\hline\noalign{\smallskip}
~$\lambda$(\AA) &  ion  & I(obs.)  &  I(corr.)\\
\noalign{\smallskip}
\hline\noalign{\smallskip}

 4686&      HeII&       30&         33&\\
 4861&       HI&       100&        100&\\
 4959&     [OIII]&     430&        400&\\
 5007&     [OIII]&    1300&       1200&\\
 5876&      HeI&        16&         10&\\
 6560&      HeII&       10:&         4:&\\
 6563&       HI&       665&        285&\\
 6584&     [NII]&       24&         10&\\
 7135&     [ArIII]&     22&          7&\\

\noalign{\smallskip}
\hline
\end{tabular}
\end{flushleft}

\end{table}
The close resemblance of the H$\alpha$ and [OIII] emission structure and the 
weakness (or the absence) of low excitation lines (in particular, [NII] at 
$\lambda$$\lambda$6548 and 6584 \AA\/ and [SII] at $\lambda$$\lambda$6717 and 
6731 \AA\/) indicate that NGC 1501 is an optically thin, density bounded PN.

\section{The expansion velocity field}

The peak separation, measured at the centre of each
nebular emission, gives the expansion velocities contained in Table 2, 
where ions are put in order of increasing ionization potential. 
For all ions (but not HI) the estimated 
error in the expansion velocity is essentially anti-correlated to the emission 
intensity; the large uncertainty in H$\alpha$ comes from the line broadening 
due to thermal motions.

\begin{table}
\caption{Expansion velocities}
\begin{flushleft}
\begin{tabular}{lcrcc}
\hline\noalign{\smallskip}
~~$\lambda$  & ion &  I.P.  &      $2\,v_{\rm exp}$ \\
~(\AA)      &     &   (eV) &      (km s$^{-1}$) \\
\noalign{\smallskip}
\hline\noalign{\smallskip}
6563&        HI&      13.6&     80 $\pm$3&\\
6584&       [NII]&    14.5&     84 $\pm$3&\\
5876&        HeI&     24.6&     82 $\pm$3&\\
7135&       [ArIII]&  27.6&     82 $\pm$3&\\   
5007&       [OIII]&   35.1&     81 $\pm$2&\\
4686&        HeII&    54.4&     79 $\pm$3&\\
\noalign{\smallskip}
\hline
\end{tabular}
\end{flushleft}

\end{table}

Previous expansion velocity measurements in the nebula go back to Robinson  et
al. (1982, $2\,v_{\rm exp}\mbox{[OIII])}=78$ $\mbox{km s}^{-1}$) and to Sabbadin 
and Hamzaoglu (1982a,\\  
$2\,v_{\rm exp} ({\rm H}\alpha)= 2\,v_{\rm exp}([OIII])=76$ $\mbox{ km s}^{-1}$).

The very low (if any) expansion gradient present in NGC 1501  confirms that 
stratification effects are negligible and suggests that the  main nebular 
emission occurs in a narrow shell (note, in particular, the large H$\alpha$ 
expansion velocity, as large as for the other lines).

More precise informations can be obtained from the detailed analysis of the  
expansion  velocity field at the four position angles. These are shown in  Figure
3 for H$\alpha$; the velocity and intensity distributions derived  for
$\lambda$5007  \AA\/ of  [OIII] coincide with the H$\alpha$ ones; though the
[NII] line at  $\lambda$6584 \AA\/ is too weak for an accurate study,  its
intensity and velocity  trends appear very similar to those observed in H$\alpha$
and [OIII].   

The H$\alpha$ radial velocity distributions of Fig. 3 present the classical 
bowed shape 
expected of an expanding shell, but they are so inhomogeneous and distorted 
that a simple model (like a triaxial ellipsoid) is decidedly 
inadequate to fit all the data. Note, in particular:

- the comparable nebular extent at P.A.=10$^o$ (perpendicular to the apparent 
major axis) and at P.A.=55$^o$ (intermediate direction);

- the noticeable differences in both the expansion velocity field and 
the nebular 
structure at P.A.=55$^o$ and 145$^o$ (being these two position angles 
symmetrically arranged with respect to the major axis, we would 
expect almost mirror structures for a simple rotational figure);

- the similarity of both the velocity field and the intensity distribution 
at P.A.=10$^o$ and 145$^o$ (suggesting that one of the axes of 
symmetry is projected in P.A.$\simeq$170$^o$). 

Moreover, the imagery of the nebula (taken a few months 
after the spectroscopic material) stands out a further complication: the 
direction 
of the apparent major axis of NGC 1501 seems closer to 110$^o$ than to 
100$^o$ (the P.A. we used), or to 120$^o$ and 98$^o$ (the P.A. given by 
Pease, 1917, and Curtis, 1918, respectively).

To make easier the interpretation of the velocity maps of Fig. 3 
in 
terms of an acceptable model, we believe convenient the 
introduction of 
a parting line at an apparent distance of about 15'' from the central star, 
separating the "low latitude" ionized gas from the "high latitude" one. 

\begin{figure*}
\resizebox{18cm}{!}{\includegraphics*[angle=-90]{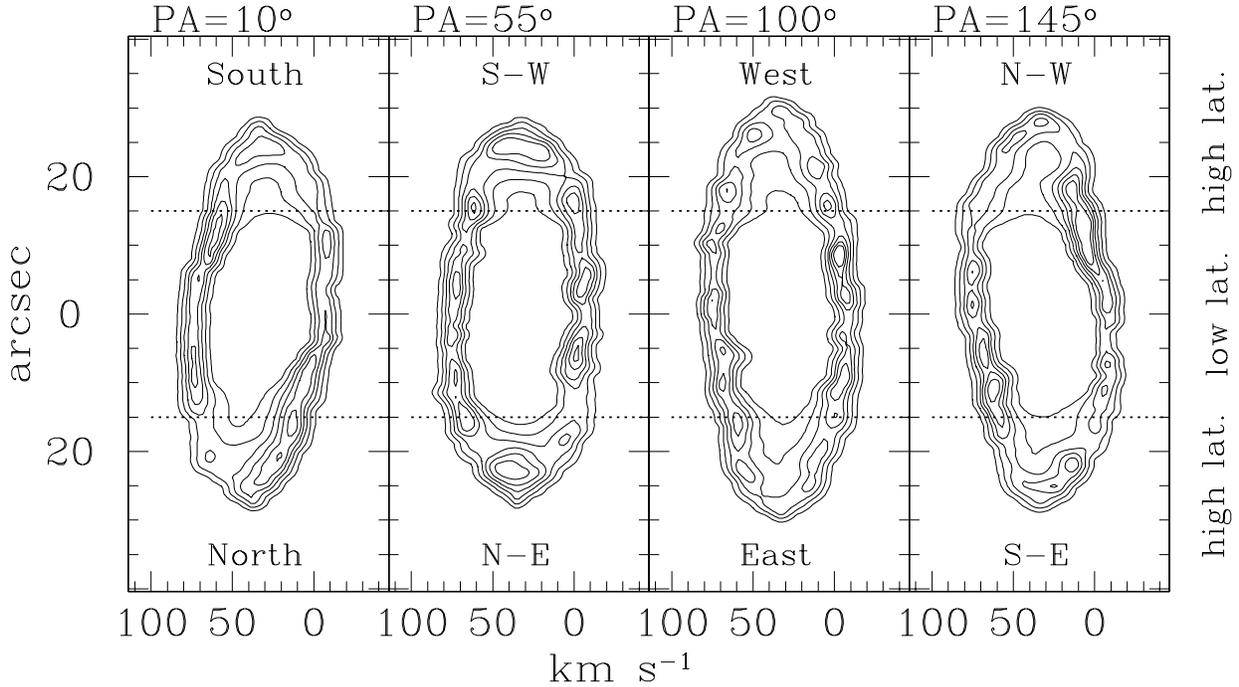}}
\caption{H$\alpha$ position-velocity contour maps observed in NGC 1501 at 
P.A.=10$^o$ 
(almost perpendicular to the direction of the apparent major axis), 
P.A.=55$^o$, P.A.=100$^o$ 
(close to the direction of the apparent major axis) and 
P.A.=145$^o$. "Low" and "high latitude" nebular regions, as introduced in the 
text, are indicated.}\label{figura2}
\end{figure*}

The ``low latitude'' velocities are quite regular elliptical arcs (i.e. 
projections of a triaxial ellipsoid), whose tilt 
is maximum at P.A.=10$^o$ and P.A.=145$^o$,  and minimum (but 
not null) at 
P.A.=100$^o$ (i.e. close to the apparent major axis). To be noticed that the
un-tilted velocity ellipse occurs at 
P.A.$\simeq$80$^o$ 
(corresponding to the line of nodes; this agrees with the foregoing 
indication that one of the axes of symmetry is projected at 
P.A.$\simeq$170$^o$). 
The intensity distribution appears 
knotty and 
irregular at P.A.= 55$^o$ and 100$^o$, while at P.A.=10$^o$ and 145$^o$ it 
presents two symmetric, extended condensations (as normally observed in 
bipolar PNe). 

The ``high latitude'' velocity maps of Fig. 3 are characterized by an 
extreme 
variability: they are faint, extended and asymmetric 
at P.A.=10$^o$ and 145$^o$, faint, extended and quite symmetric at 
P.A.=100$^o$, 
knotty, bright, symmetric and small at P.A.=55$^o$.
The general impression is that they represent hemispheric bubbles protruding 
from the ellipsoidal region of NGC 1501; in some directions 
(as in the East sector 
of P.A.=100$^o$) multiple structures appear, i.e. an external, smaller dome 
overlaps the internal, larger one.

A further, remarkable feature of the ``high latitude'' zones in Fig. 3 
is the 
internal, weak, diffuse emission present at all the four position angles, 
indicating that a low density ionized gas completely fills these nebular 
regions.

In order to decipher the complex gas motion observed in NGC 1501, a detailed 
analysis of the nebular physical conditions will be performed in the next 
Section.

\section{Physical conditions}

\subsection{Electron temperature (Te)}

In absence of diagnostic line intensity ratios (e.g. 4363/5007 \AA\/ of [OIII] 
and 5755/6584 \AA\/ of [NII]), the electron temperature of the ionized gas
was derived by comparing the H$\alpha$ and [OIII] emission line profiles; the 
basic assumption is that, for a given $Te$, the thermal motion in H is four 
times larger that in O (the former element being sixteen times lighter that 
the 
second).

{\bf CAVEAT}:{\em This method can be rightly applied to PNe only if the H$^+$ 
and 
O$^{++}$ layers do coincide (as in NGC 1501, which is an optically thin, high 
excitation PN). In many cases, the presence of large 
stratification effects invalidates the results, since H$^+$ and O$^{++}$ are 
emitted in separated layers expanding in different ways (due to the 
presence of a gradient in the velocity field).
Similarly, the H$^+$ and N$^+$ line profile comparison fails in medium and 
high excitation PNe.}

The full width at half maximum (W) of an emission line is the 
convolution 
of different components:

\begin{equation}
W_{\rm total}^2=W_{\rm instr.}^2+W_{\rm thermal}^2+W_{\rm turbulence}^2+
W_{\rm gradient}^2   
\end{equation}

where the last (too frequently neglected) term takes into account the 
existence of a radial gradient in the expansion velocity field.  

A further broadening factor concerns H$\alpha$: its 
seven fine 
structure components can be modeled as the sum of two equal Gaussians 
separated by 0.14 \AA\/ (Dyson and Meaburn, 1971, and Dopita, 1972).

Since in NGC 1501 the H$^+$ and O$^{++}$ layers do coincide, we have:

\begin{equation} T{\rm e} = f[W(H\alpha)_{total}^2 - W([OIII])_{total}^2]     
\end{equation}

A mean value of $T{\rm e}=11500\pm500 \mbox{ K}$ is derived by analysing both the
blue-shifted  and red-shifted H$\alpha$ and [OIII] profiles in the central
region of the lines; the  spread in $T$e, only partially due to instrumental +
measurement  uncertainties, testifies the existence of electron temperature
variations  within the ionized gas, although no clear correlation was found with
position  and/or flux.

The only previous $T$e determination in this nebula is reported by 
Stanghellini et al.
(1994), who obtained $T$e=10700 K from the 4363/5007 \AA\/ [OIII] intensity 
ratio. 
Moreover, some statistical works on a large sample of PNe (e.g. Cahn et al., 
1992, and Phillips, 1998) adopt $T$e=15100 K, based on the strength of the 
HeII $\lambda$4686 \AA\/ line and Kaler's, 1986, calibration.

Having obtained $T$e, the value of ($W_{turb.}^2 + 
W_{grad.}^2$) can 
be inferred from relation (1); since in NGC 1501 the expansion velocity 
gradient is small (see Table 2), it can be neglected, allowing us 
to quantify 
turbulent motions in the ionized gas. We obtain for $W_{\rm turbulence}$ a 
mean value of 18
$\mbox{km s}^{-1}$ from both the H$\alpha$ and [OIII] profiles, but small scale 
fluctuations (up to $10 \mbox{ km s}^{-1}$) are present. 

A comparison with previous 
data reported in the literature for PNe gives poor 
results, because of our "caveat", of the scarce bibliography and of the 
different reduction methods. 
We recall the analysis 
performed by Gesicki et al. (1998) on seven PNe, indicating that the highest 
turbulent motions (15 km s$^{-1}$) occur in M 3-15, a nebula powered by a WC 
4-6 star. Although this result is weakened by the evidence that M 3-15 is 
ionization bounded (at least in some directions), we agree with the statement 
of these authors: "It is possible that nebulae with [WC] central stars have 
less regular velocity fields than the other PNe... The assumed very high 
turbulence may only be an approximation to a more complicated situation with 
strong velocity variations in radial direction". 

\subsection{Electron density (Ne)}

Also $N$e diagnostic line ratios 
(e.g. 6717/ 6731 \AA\/ 
of 
[SII], 3726/3729 \AA\/ of [OII], 4711/4740 \AA\/ of [ArIV] and 5517/5537 \AA\/ 
of [ClIII]) are absent in our echellograms of NGC 1501. In order to 
derive the electron density, we will proceed in three steps:

a) determination of the relative $N$e radial trend
in the ``zero-velocity pixel column'' (as defined by Sabbadin et al., 2000);

b) transformation to absolute $N$e values by means of a suitable calibrator;

c) extension of the absolute $N$e determination to the whole nebula's slice 
covered by the slit.

Points a) and b) will be discussed here and point c) in the next Section, 
dedicated to tomography.

\begin{figure}
\resizebox{\hsize}{!}{\includegraphics*{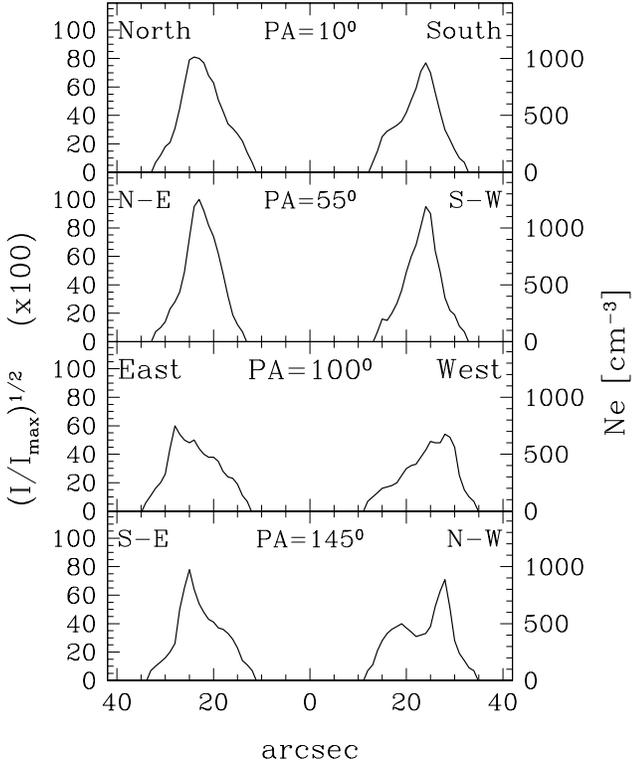}}
\caption{Relative (left ordinate scale) and absolute (right ordinate scale) 
radial density trends obtained from the H$\alpha$ zero-velocity pixel columns.
Note 
the steep outwards profile and the broad inwards tail present at all the 
four position angles (less evident in P.A.=55$^o$).} \label{fig 3}
\end{figure}

Let's consider a long-slit, spatially resolved, high resolution spectrum of a 
typical planetary nebula expanding at $v_{\rm exp}$. Following Sabbadin et al.
(2000), the zero-velocity pixel column of an emission line represents the slice
of  nebula  centred in the plane of the sky crossing the exciting star and having
a  depth along the line of sight $t=r\Delta v/v_{\rm exp}$, where $r$ is the 
nebular radius and  $\Delta v$ is  the pixel spectral resolution. As an example,
in the velocity maps shown in Fig. 3, the observed H$\alpha$ zero-velocity column
at each position angle is represented by a vertical strip centred at 
$+36 \mbox{ km s}^{-1}$ (systemic radial velocity of NGC 1501) and 
$9.0 \mbox{ km s}^{-1}$ wide.    

At every position along the slit, the intensity in the zero-velocity column is 
proportional to $N{\rm e}N{\rm i}$; in the case of complete ionization and 
$T$e=constant, I$\propto N{\rm e}^2$ 
(we implicitly assume the constancy of the local filling factor, 
$\epsilon_l$, within the slice of nebula identified by the zero-velocity 
pixel 
column; $\epsilon_l$ is the fraction of the local nebular volume 
which 
is actually filled by matter with density $N$e; the local nebular volume is 
given by: (pixel area)$\times t$, where $t=r \Delta v/v_{\rm exp}$).

In short: the zero-velocity pixel column, isolating a slice of nebula
 unaffected by the expansion velocity field, establishes  a direct
 link between the intensity profile and the ionic and electron
 density distributions, thus allowing us the detailed
analysis of the radial gas structure (and ionization) in
the expanding nebula.

The H$\alpha$ zero-velocity column distributions, corrected for contamination 
by the
adjacent spectral columns and for seeing and guiding uncertainties (for 
details, 
see Sabbadin et al., 2000), are shown in Figure 4; they 
are normalized to the strongest 
intensity (i.e. the peak in the N-E sector of P.A.=55$^o$). 

The radial density profile of NGC 1501 results quite complex; the 
main features, common at all the four position angles,  are the 
following:

- the outermost parts correspond to the faint, roundish halo detected in our 
R image (representing the vestiges of the photospheric material ejected 
in the AGB phase);

- the density peaks are located in the external nebular regions (close to the 
edge of the bright disk visible in Figs. 1 and 2); their distribution 
confirms that 
the main component of NGC 1501 is a shell of moderate thickness;

- the density peaks show steep outwards profiles;

- inwards tails of different shapes are present at all the four 
position angles; their extent seems anti-correlated to the height of the 
density peak: small at P.A.=55$^o$, intermediate at P.A.=10$^o$, broad at 
P.A.=145$^o$ (note the detached structure in the N-W sector) 
and at P.A.=100$^o$ (note the jagged profile).

Very similar radial density trends are obtained by analysing the [OIII]  
zero-velocity pixel columns.  

In order to scale to absolute values the relative electron density profiles 
shown in Fig. 4, we will utilize the direct intensity-$N{\rm e}$ (cm$^{-3}$) 
relation 
given 
by the H$\alpha$ zero-velocity pixel column of NGC 40 (observed in the same 
nights 
and with 
the same instrumental setup used for NGC 1501).

We recall that the [SII] red doublet is quite strong in the low excitation 
PN NGC 40 (Aller and Epps, 1976, and Clegg et al., 1983), thus an accurate 
determination of $N$e can 
be obtained from the diagnostic ratio 6717/6731 \AA\/ (Sabbadin et al., 2000).

From the general expression:

\begin{equation}
4\pi D^2 F(H\alpha)= h \nu \int_{0}^{R}\alpha_{3,2}\, N(H^+)N{\rm e}\,\epsilon 4
\pi r^2 dr    
\end{equation}

we obtain the following relation between the surface brightnesses of the 
zero-velocity pixel columns of NGC 1501 and NGC 40:

\begin{equation}
\frac{N{\rm e}_1}{N{\rm e}_2} = 
\biggl(\frac{I_1 v_{\rm exp\,1} r''_2 D_2 \epsilon_{l,2}} 
{I_2 v_{\rm exp\,2}  r''_1 D_1 \epsilon_{l,1}}\biggr)^{0.5} 
\biggl(\frac{T{\rm e}_2}{T{\rm e}_1}\biggr)^{-0.44} 
\end{equation}

where the suffixes (1) and (2) refer to NGC 1501 and to NGC 40, respectively, 
$I$ is the H$\alpha$ 
intensity 
in a pixel of the zero-velocity column (corrected for 
interstellar extinction), $r''$ is the apparent nebular radius, $D$ is the 
distance and $\epsilon_l$ is the local filling factor (as previously 
defined). For both nebulae we assumed $N{\rm e}=1.1 N({\rm H}^+$).

The calibration through relation (4) gives for NGC 1501 the absolute radial 
density 
profiles shown in Fig. 4, right ordinate scale.
The following parameters (some taken from the literature) were adopted:

NGC 40: c(H$\beta$)=0.60, $T$e=8000 K, $v_{\rm exp}$=25 km s$^{-1}$, $r''$=20'', 
$D$=1.1 Kpc.

NGC 1501: c(H$\beta$)=1.05, $T$e=11500 K, $v_{\rm exp}$=40 km s$^{-1}$, $r''$=28'',
 $D$=1.3 Kpc.

Moreover, we assumed $\epsilon_l(\mbox{NGC 40})=\epsilon_l(\mbox{NGC 1501})=1$; 
in general, this appears to  be a reasonable condition; some 
particular cases will be discussed later-on.

The estimated inaccuracy in the $N$e scaling factor is $\pm$20\%, 
mainly due to  
our poor knowledge of the PNe distances; fortunately, in relation (4) the 
electron density has a low dependance on the distance.

We wish to point out that the foregoing method [i.e. the use of a PN (NGC 40) 
as a density 
calibrator for an other PN (NGC 1501)] represents a procedural {\em escamotage
}: we were forced to adopt it since the night sky during the observational 
run was stable only at intervals. A more elegant analysis, based on the 
``absolute'' surface brightness [i.e. formula (3)] will be presented and 
discussed in a future paper dedicated to the double envelope PN NGC 2022.

\section{Tomography}

\begin{figure*}
\resizebox{18cm}{!}{\includegraphics*{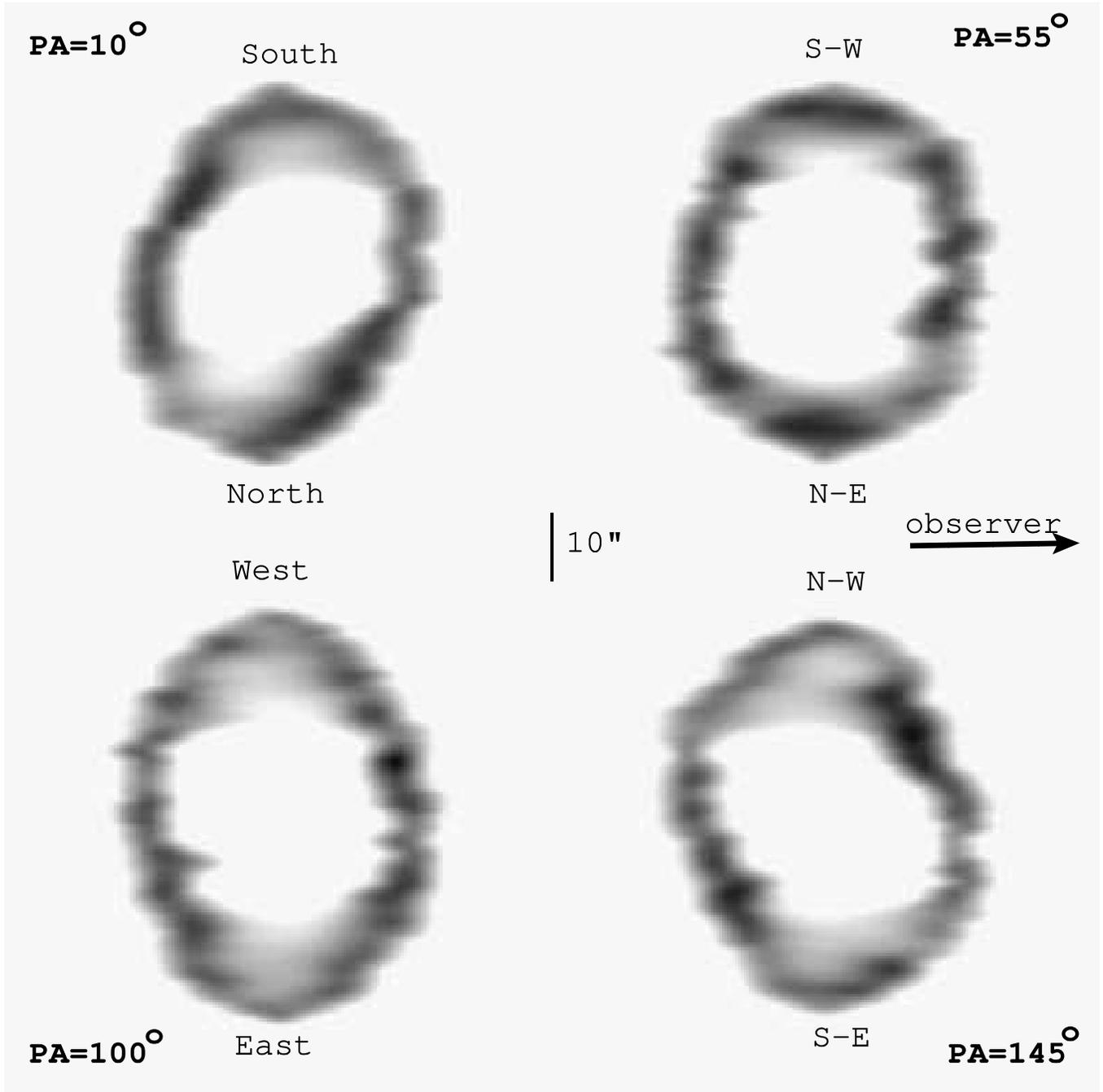}}
\caption{Spatial reconstruction of the $N$e structure of 
NGC 1501 derived at P.A.=10$^o$ (almost perpendicular to the direction of the 
apparent major axis), P.A.=55$^o$, 
P.A.=100$^o$ (close to the direction of the apparent major axis) and 
P.A.=145$^o$. The minimum density shown is $N$e=200 cm$^{-3}$, while the 
maximum 
density ($N$e=1380 cm$^{-3}$) is reached by the 
approaching gas located in P.A.=145$^o$ (N-W sector), at an apparent distance 
of 13'' from the central 
star. Being NGC 1501 an optically thin, high excitation PN, the $N({\rm H}^+)$ 
and $N({\rm O}^{++})$ tomographic maps coincide with the $N$e ones (but for a 
scaling factor of 1.1 and $3.67\times10^3$, respectively).} \label{figura5}
\end{figure*}

To derive the de-projected expansion velocity field and the real gas 
distribution within the slice of nebula covered by the spectroscopic slit, 
we will follow the original 
method described by Sabbadin (1984) and Sabbadin et al. (1985), and  recently 
applied by Sabbadin et al. (2000) to the low excitation PN NGC 40.

The basic consideration is that  the 
position, radial thickness and density of each emitting region in an extended, 
expanding object can be obtained from the velocity, FWHM and flux, 
respectively.

In the case of NGC 1501, the limited number of emissions detected and the 
total absence of stratification effects across the nebula force us to use a 
linear position-speed relation. Moreover, from the angular extent of the 
ellipses fitting the "low latitude" nebular regions in Fig. 3, we will adopt 
$r$$_1$ = 
20", where $r$$_1$ is the nebular radius in the radial direction at the 
apparent position of the central star (for details, see Sabbadin, 1984).
Finally, an electron temperature of 11500 K and a turbulence of 18 km s$^{-1}$ 
(both constant over the nebula) are assumed.

The adoption of other (reasonable) sets of parameters doesn't substantially 
modify the results here obtained.

Our tomographic analysis of NGC 1501 mainly concerns the electron density; 
the $N$e structure of the nebula at the 
four position angles is shown in Figure 5.
According to the discussion above, the $N({\rm H}^+$) tomographic 
maps coincide with the $N$e ones (but for a scaling factor of 1.1). The same 
situation occurs for $N({\rm O}^{++})$, but in this case the scaling factor is 
$3.67\times10^3$; in fact, the $\lambda$5007 \AA\/ intensity reported in 
Table 1 
gives $N({\rm O}^{++})/N({\rm H}^+) = 3.0(\pm0.4)\times10^{-4}$ (see Aller and 
Czyzak, 
1983 and Aller and Keyes, 1987).
Moreover, the ionization structure of NGC 1501 suggests that most oxygen is 
doubly ionized, so that $N({\rm O})/N({\rm H}) \simeq N({\rm O}^{++})/N({\rm H}^+)$; to be 
noticed that Stanghellini et al. (1994) report for this nebula 
$N({\rm O}^{++})/ N({\rm H}^+) = 3.72\times10^{-4}$ and $N({\rm O})/N({\rm H}) = 
5.7\times10^{-4}$.  

The only previous $N$e determination in NGC 1501 dates back to Aller and Epps 
(1976), who observed at low spectral resolution a small region centred 20'' 
from the 
star (in P.A.=125$^o$); they derived $I(6717)/I(6731)=0.78$, corresponding to 
$N$e=1200 cm$^{-3}$ (for $T$e=11500 K). Their direction is intermediate 
between 
our P.A.=100$^o$ and P.A. =145$^o$; at a distance of 20'' from the central 
star 
we obtain the following density peaks:

$N$e(blue shifted)=900 cm$^{-3}$ and $N$e(red shifted)=1000 cm$^{-3}$ in 
P.A.=100$^o$;

$N$e(blue shifted)=700 cm$^{-3}$ and $N$e(red shifted)=900 cm$^{-3}$ in 
P.A.=145$^o$.
 
Unfortunately, a more detailed comparison appears hazardous, due to the 
differences in the
spectroscopic resolution and reduction procedure, and, in particular, to the 
presence of small scale density fluctuations within the nebula.

The de-projected expansion velocities of NGC 1501 (directly obtainable 
from Fig. 5, given the linear position-speed relation used) span in 
the range 38($\pm$2) to 
55($\pm$2) km s$^{-1}$. The slowest motions occur in the densest regions at 
P.A.=10$^o$ and P.A.=145$^o$; the combination (high density + low expansion 
velocity) here observed suggests that the minor axis of the central ellipsoid 
is projected 
at P.A.$\simeq$170$^o$ (in agreement with the indications given in Section 3).
The largest expansion velocities [55($\pm$2) km s$^{-1}$] correspond to
the high latitude, 
untilted, hemispheric bubbles at P.A.=100$^o$; in this case, the 
combination 
(low density + high velocity) suggests that we are (almost) observing the 
projection of the major axis of the central figure. Finally, from geometrical 
considerations, the projection of the intermediate axis can be put at 
P.A.$\simeq$30$^o$.

Figure 5 summarizes most of the observational results given in the 
previous 
sections, confirming the composite structure of the nebula: NGC 1501 is an 
ellipsoid of moderate ellipticity, deformed by a pair of large lobes along 
both the major 
and intermediate axes and by a number of minor bumps scattered on the 
whole nebular surface (in a few cases, see for instance the big ``ear'' in the 
N-W sector of P.A.=145$^o$, the dimensions of these ``minor bumps'' appear 
comparable to those of the lobes related to the axes of the central ellipsoid).

In Fig. 5, the absence of the broad, inwards tail at low latitude is
probably due to instrumental limitations:
intuitively, our echellograms of an extended PN like NGC 1501 have 
a ``spatial'' resolution along the 
slit which is better than the ``spectral'' resolution along the dispersion. 
Due to 
projection effects, the "low latitude" zones (dominated by the expansion 
velocity field) have a "spectral" resolution, the "zero-velocity pixel 
column" has a "spatial" 
resolution (being unaffected by the expansion velocity field) and the "high 
latitude" zones an hybrid resolution. Clearly, this gradual 
variation in resolution conditions the tomographic reconstruction.
In other words: we cannot exclude the presence of a low density, inwards 
tail also at low latitude. Only deep observations at much
higher spectral resolution could solve the question.

Having said this,  we believe that Fig. 5 adequately reproduces the 
true matter distribution in NGC 1501,
 and that the faint, inwards emissions 
visible at high latitude do represent the trace of the original ellipsoidal 
structure (note, in particular, the sharp radial profile in the high latitude 
zone at P.A.=55$^o$ and the detached structure in the N-W sector of 
P.A.=145$^o$, fitting the ellipsoid projection). 
This implies that some accelerating 
agent partly swept-up the lower density regions of the triaxial ellipsoid,
causing both the extended, hemispheric bubbles and the broad, inwards density 
tails. The final result of this acceleration is constituted by the spatial 
structure illustrated in the next Section.

\section{Spatial model}

An opaque sketch of the resulting 3-D model for NGC 1501 
is given in Figure 6, superimposed to the R frame of the nebula. 
The outermost 
contour 
of the model is enhanced; moreover, Fig. 6 contains:

- the directions (A, B and C) of the semi-axes a, b and c of the central 
ellipsoid;

- the large lobes associated to the major and intermediate axes;

- some ``secondary'' bubbles identified in the spectra and/or in the imagery;
due to projection effects, only those at (or near) the nebular edge are
detectable;

- the extent of the faint, roundish, external envelope visible in 
the H$\alpha$ 
image.  

\begin {figure}
\resizebox{\hsize}{!}{\includegraphics*{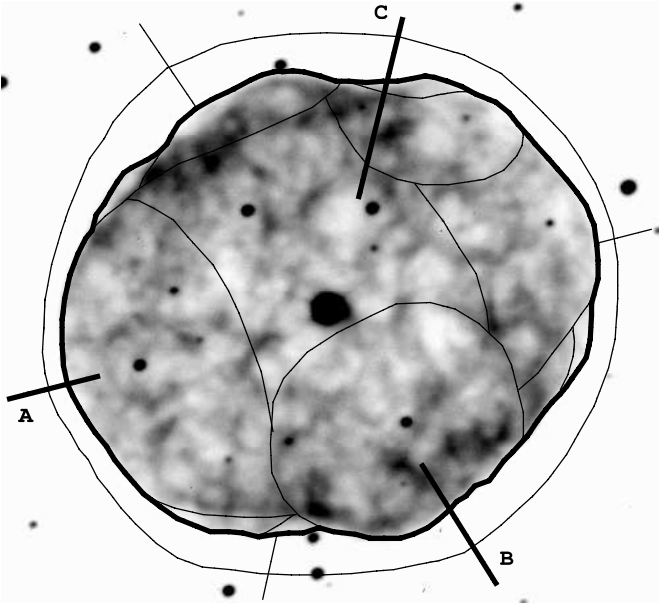}}
\caption{Opaque sketch of the resulting spatial structure of 
NGC 1501 
(superimposed to the red frame of the nebula). The model 
contains:
- the directions (A, B and C) of the semi-axes (a, b and c) of the central 
ellipsoid;
- the large lobes associated to the major and intermediate axes;
- some secondary lobes visible at (or near) the nebular edge;
- the extent of the faint, roundish envelope detected in the R image. Same 
orientation and scale as Figs. 1 and 2.} 
\label{figura 6}
\end{figure}

Although rare, the peculiar morphology of NGC 1501 is not unique amongst PNe.
A quick look at the main imagery catalogues allowed us to identify 
three more candidates: A 72 and NGC 7094 are faint, high 
excitation PNe presenting a complex filamentary structure (Manchado et al., 
1996a); IC 4642 is quite bright, at high excitation, decidedly 
tetra-lobed in both the H$\alpha$ and [OIII] images published by Schwarz et 
al. (1992).

Moreover, Manchado et al. (1996b) introduced the morphological class 
of quadrupolar PNe, containing five compact objects (M 2-46, K 3-24, M 1-75, 
M 3-28 and M 4-14); these nebulae present two pairs of lobes, each pair 
symmetric with 
respect to a different axis. A sixth candidate, NGC 6881, was added by 
Guerrero and Manchado (1998). Following these authors, a quadrupolar PN 
can be 
formed by precession of the rotation axis of the central AGB star, possibly 
in the presence of a binary companion, associated with multiple shell 
ejection. 

Finally, the general spatial structure derived for NGC 1501 (i.e. two pairs 
of bipolar lobes on axes having different orientations and intersecting at 
the central star) presents noticeable analogies with the model suggested 
by Balick and Preston (1987, their Fig. 4) for NGC 6543.

In the case of NGC 1501, the modest departures from the spherical 
symmetry 
can be explained in terms of small inhomogeneities occurred during the 
super-wind ejection (for instance, due to stellar rotation), later enhanced 
by photoionization and winds interaction (see Dwarkadas and Balick, 1998 and 
Garcia-Segura et al., 1999). The introduction of a close companion 
(as suggested for NGC 6543 by Balick and Preston, 1987, and 
Miranda and Solf, 1992), or even of a binary companion (as proposed by 
Manchado et al., 1996b, for quadrupolar PNe) appears unnecessary for NGC 
1501.

Clearly, the tetra-lobed shape is a simplification of the true spatial  
structure of this nebula; a close inspection to the observational data 
indicates 
that each macro-lobe is constituted of a heap of small components, that 
some morphological differences exist amongst the lobes associated to the 
major axis of the ellipsoid and those connected with the intermediate 
axis, and that these lobes are only roughly aligned with the 
axes of the central figure.
An extended spectroscopic coverage of the nebula is in progress, in 
order to obtain the detailed spatial structure of the 
ionized gas.

The most intriguiging characteristic of the matter distribution in NGC 1501 
is the 
presence of an inwards tail in the radial density trend; this tail, 
particularly evident in the directions of the lobes, can be the result of 
hydrodynamic processes in the nebular shell. Following Capriotti (1973; see 
also Breitschwerdt and Kahn, 1990, Kahn and Breitschwerdt, 1990, and
Garcia-Segura et al., 1999), in the first evolutionary phases, when the PN is 
still ionization bounded, Rayleigh-Taylor instabilities occur at the 
ionization front, forming a series of knots, condensations and radially 
arranged fingers (see Dyson, 1974, and Bertoldi and McKee, 1990) which 
expand slower than the ionization front.

Similar instabilities are produced also by the interaction of the fast stellar 
wind with the low velocity nebular material (Vishniac, 1994, Garcia-Segura 
and Mac Low, 1995, and Dwarkadas and Balick, 1998); in this case the optical 
thickness of the nebula is unimportant. If confirmed, the extremely large 
value of $\dot{M} $
($\log \dot{M}=-6.28$ ${\rm M}\sun\, yr^{-1}$) derived for the WC4/OVI nucleus of NGC 
1501 by Koesterke 
and Hamann (1997a) using the standard atmosphere model for Wolf-Rayet stars, 
would indicate winds interaction as the main responsible of both the radial 
density distribution and the large expansion velocity observed in this nebula.

To be noticed that, besides the dynamical effects on the nebular gas, an 
intense and lasting mass-loss of hydrogen depleted and He, C and O enriched 
photospheric material would produce enhanced ionic and 
chemical composition gradients across the nebula (as recently observed by 
Sabbadin et al., 2000 in NGC 40, a low excitation PN powered by a WC8 star).

Moreover: if winds interaction and/or Rayleigh-Taylor instabilities  
are the sources of the inwards tails detected in the density 
distribution of NGC 1501, these tails are essentially constituted of knots and 
condensations; if they survived ionization and/or heating by conduction, 
we expect $\epsilon_l(\mbox{tail})<\epsilon_l(\mbox{main shell})$, 
where $\epsilon_l$ is
the local filling factor, as previously defined. In other words: the true 
electron densities in the blobs and condensations forming the inwards tail are
larger by the factor 
$[\epsilon_l(\mbox{main shell}) /\epsilon_l(\mbox{tail})]^{0.5}$ than 
the values shown in Figs. 4 and 5, where we assumed 
$\epsilon_l(\mbox{tail})=\epsilon_l(\mbox{main shell})=1$.  

Unfortunately, our observational material is inadequate to transform the
previous speculations into quantitative results to be compared with 
the theoretical predictions. Detailed, deep studies at higher 
spatial and spectral resolution of this (overlooked so far) PN are needed 
to answer the stimulating questions here excited.

\section{Conclusions}

The radial density distribution of the ionized gas in the high 
excitation PN NGC 1501 presents density peaks up 
to 1400 cm$^{-3}$; they have steep outwards profiles and extended inwards 
tails probably originated by Rayleigh-Taylor instability and winds 
interaction.

By comparing the H$\alpha$ and [OIII] emission line profiles we derive an 
electron temperature of 11500 K and a turbulence of 18 km s$^{-1}$, but large, 
small scale fluctuations of both these quantities are present in the ionized 
gas.

The complex expansion velocity field observed in this nebula is fitted by an 
ellipsoid of moderate 
ellipticity, deformed by a pair of large lobes along both the major and 
intermediate axes and by a multitude of minor bumps scattered on the whole 
nebular surface.

The peculiar morphology of NGC 1501 can be qualitatively explained in 
terms of interaction of the nebular gas with the intense and fast wind from 
the WC4/OVI central star.

The exquisitely observational nature of this paper exempts us for hunting 
sophisticate evolutionary models; since our race is finished, we strecth the 
baton to someone else (hoping to see a ``modellist'' ready to catch 
it and to put on a winning burst of speed).

\begin{acknowledgements}
We are grateful to the referee (R. Tylenda) for his critical comments, which 
improved the paper considerably.
We thank the night 
assistants and the whole technical staff of the Astronomical Observatory 
of Asiago at Cima Ekar (Italy) for their
competence and patience.

This paper is partially based on observations made with the Italian Telescopio
 Nazionale Galileo (TNG) operated on the island of La Palma by the Centro 
Galileo Galilei of the CNAA (Consorzio Nazionale per l'Astronomia e 
l'Astrofisica) at the Spanish Observatorio del Roque de los Muchachos of the 
Instituto de Astrofisica de Canarias.  
\end{acknowledgements}

\bibliographystyle{astron}
\bibliography{/home/enrico/testi/bibliografia}

\end{document}